\documentclass[12pt]{article}

\usepackage{amsfonts,amscd,amssymb}
\usepackage[matrix,arrow,frame,curve]{xy}
\CompileMatrices

\begin{document}
\def\IB{\relax\hbox{$\inbar\kern-.3em{\rm B}$}}
\def\IC{\relax\hbox{$\inbar\kern-.3em{\rm C}$}}
\def\ID{\relax\hbox{$\inbar\kern-.3em{\rm D}$}}
\def\IE{\relax\hbox{$\inbar\kern-.3em{\rm E}$}}
\def\IF{\relax\hbox{$\inbar\kern-.3em{\rm F}$}}
\def\IG{\relax\hbox{$\inbar\kern-.3em{\rm G}$}}
\def\IGa{\relax\hbox{${\rm I}\kern-.18em\Gamma$}}
\def\IH{\relax{\rm I\kern-.18em H}}
\def\IK{\relax{\rm I\kern-.18em K}}
\def\IL $N=2${\relax{\rm I\kern-.18em L}}
\def\IP{\relax{\rm I\kern-.18em P}}
\def\IR{\relax{\rm I\kern-.18em R}}

\def\IZ {{\mathcal{I}_Z}}

\def\ocom{\relax{\raise1.2pt\hbox{$\otimes$}\kern-.5em\lower.1pt\hbox{,}}\,}
\def\CA {{\cal A}}
\def\CB {{\cal B}}
\def\CC {{\cal C}}
\def\CD {{\cal D}}
\def\CE {{\cal E}}
\def\CF {{\cal F}}
\def\CG {{\cal G}}
\def\CH {{\cal H}}
\def\CI {{\cal I}}
\def\CJ {{\cal J}}
\def\CK {{\cal K}}
\def\CL {{\cal L}}
\def\CM {{\cal M}}
\def\CN {{\cal N}}
\def\CO {{\cal O}}
\def\CP {{\cal P}}
\def\CQ {{\cal Q}}
\def\CR {{\cal R}}
\def\CS {{\cal S}}
\def\CT {{\cal T}}
\def\CU {{\cal U}}
\def\CV {{\cal V}}
\def\CW {{\cal W}}
\def\CX {{\cal X}}
\def\CY {{\cal Y}}
\def\CZ {{\cal Z}}

\def\H {{\bf{H}}}


\def\wwp{{\wp}_{\tau}(q)}
\def\gpg{g^{-1} \p g}
\def\pgp{\pb g g^{-1}}
\def\p{\partial}
\def\pb{\bar{\partial}}
\def\ib{{\bar i}}
\def\jb{{\bar j}}

\def\ub{{\bar{u}}}
\def\wb {\bar{w}}
\def\zb {\bar{z}}

\def\codim{{\mathop{\rm codim}}}
\def\cok{{\rm cok}}
\def\rank{{\rm rank}}
\def\coker{{\mathop {\rm coker}}}
\def\diff{{\rm diff}}
\def\Diff{{\rm Diff}}
\def\ib{{\bar i}}
\def\Tr{\rm Tr}
\def\Id{{\rm Id}}
\def\vol{{\rm vol}}
\def\Vol{{\rm Vol}}
\font\manual=manfnt \def\dbend{\lower3.5pt\hbox{\manual\char127}}
\def\danger#1{\smallskip\noindent\rlap\dbend%
\indent{\bf #1}\par\vskip-1.5pt\nobreak}
\def\c{\cdot}
\def\half {{1\over 2}}
\def\ch{{\rm ch}}
\def\Det{{\rm Det}}
\def\DET{{\rm DET}}
\def\Hom{{\rm Hom}}
\def\imp{$\Rightarrow$}
\def\IX{{\bf X}}
\def\neib{neighborhood}
\def\inbar{\,\vrule height1.5ex width.4pt depth0pt}

\def\Lie{{\rm Lie}}
\def\lieg{{\underline{\bf g}}}
\def\lieh{{\underline{\bf h}}}
\def\liet{{\underline{\bf t}}}
\def\liek{{\underline{\bf k}}}
\def\sdtimes{\mathbin{\hbox{\hskip2pt\vrule height 4.1pt depth -.3pt width
.25pt
\hskip-2pt$\times$}}}
\def\clb#1#2#3#4#5#6{\pmatrix{#1 & #2 & #3\cr
#4 & #5 & #6\cr} }

\def\QATOPD#1#2#3#4{{#3 \atopwithdelims#1#2 #4}}
\def\stackunder#1#2{\mathrel{\mathop{#2}\limits_{#1}}}
\def\bea{\begin{eqnarray}}
\def\eea{\end{eqnarray}}
\def\nn{\nonumber}
\newcommand{\Op}{\bigoplus}
\newcommand{\op}{\oplus}
\newcommand{\we}{\wedge}

   \newcommand{\be}{\begin{equation}}
   \newcommand{\ee}{\end{equation}}
   \newcommand{\ba}{\hspace*{-5pt}\begin{array}}
   \newcommand{\ea}{\end{array}}
   \newcommand{\ds}{\displaystyle}

 \newcommand{\ga}{\gamma}
 \newcommand{\Ga}{\Gamma}
 \newcommand{\de}{\delta}
 \newcommand{\eps}{\epsilon}
 \newcommand{\De}{\Delta}
 \newcommand{\om}{\omega}
 \newcommand{\Om}{\Omega}
 \newcommand{\Na}{\Nabla}

   \newcommand{\rt}{\rightarrow}
   \newcommand{\llrt}{\Longleftrightarrow}
   \newcommand{\lrt}{\Longrightarrow}
   \newcommand{\la}{\lambda}

\def\baselinestretch{1.5}
\def\beq{\begin{equation}}
\def\eeq{\end{equation}}
\def\ba{\beq\new\begin{array}{c}}
\def\ea{\end{array}\eeq}
\def\be{\ba}
\def\ee{\ea}
\def\stackreb#1#2{\mathrel{\mathop{#2}\limits_{#1}}}
\def\Tr{{\rm Tr}}
\def\res{{\rm res}}
\def\f{1\over}
\parskip=0.4em
\makeatletter
\newdimen\normalarrayskip              
\newdimen\minarrayskip                 
\normalarrayskip\baselineskip
\minarrayskip\jot
\newif\ifold             \oldtrue            \def\new{\oldfalse}
\def\arraymode{\ifold\relax\else\displaystyle\fi} 
\def\eqnumphantom{\phantom{(\theequation)}}     
\def\@arrayskip{\ifold\baselineskip\z@\lineskip\z@
     \else
     \baselineskip\minarrayskip\lineskip2\minarrayskip\fi}
\def\@arrayclassz{\ifcase \@lastchclass \@acolampacol \or
\@ampacol \or \or \or \@addamp \or
   \@acolampacol \or \@firstampfalse \@acol \fi
\edef\@preamble{\@preamble
  \ifcase \@chnum
     \hfil$\relax\arraymode\@sharp$\hfil
     \or $\relax\arraymode\@sharp$\hfil
     \or \hfil$\relax\arraymode\@sharp$\fi}}
\def\@array[#1]#2{\setbox\@arstrutbox=\hbox{\vrule
     height\arraystretch \ht\strutbox
     depth\arraystretch \dp\strutbox
     width\z@}\@mkpream{#2}\edef\@preamble{\halign
\noexpand\@halignto
\bgroup \tabskip\z@ \@arstrut \@preamble \tabskip\z@ \cr}%
\let\@startpbox\@@startpbox \let\@endpbox\@@endpbox
  \if #1t\vtop \else \if#1b\vbox \else \vcenter \fi\fi
  \bgroup \let\par\relax
  \let\@sharp##\let\protect\relax
  \@arrayskip\@preamble}
%
%
%
%
\def\eqnarray{\stepcounter{equation}%
              \let\@currentlabel=\theequation
              \global\@eqnswtrue
              \global\@eqcnt\z@
              \tabskip\@centering
              \let\\=\@eqncr
              $$%
 \halign to \displaywidth\bgroup
    \eqnumphantom\@eqnsel\hskip\@centering
    $\displaystyle \tabskip\z@ {##}$%
    \global\@eqcnt\@ne \hskip 2\arraycolsep
         $\displaystyle\arraymode{##}$\hfil
    \global\@eqcnt\tw@ \hskip 2\arraycolsep
         $\displaystyle\tabskip\z@{##}$\hfil
         \tabskip\@centering
    &{##}\tabskip\z@\cr}

\begin{center}
\hfill ITEP/TH-02/01\\
\vspace{0.3in}
\setcounter{footnote}0
{\LARGE\bf Dualities in integrable systems: geometrical aspects}

\bigskip

{\Large A. Gorsky,\footnote{ permanent address; ITEP, 25,
B.Cheryomushkinskaya,117259, Moscow, Russia} V. Rubtsov}

\bigskip

ITEP, 25, B.Cheryomushkinskaya,117259, Moscow, Russia and\\
Departement de Mathematics, University d'Angers,\\
2, Bd Lavoisier, 49045, Angers, Cedex 01, France\\
\end{center}

\bigskip

\begin{abstract}
We discuss geometrical aspects of different dualities in the
integrable systems of the Hitchin type and its various
generalizations. It is shown that T duality known in the string
theory context is related to the separation of variables procedure
in dynamical system. We argue that there are analogues of S duality
as well as  mirror symmetry in the many-body systems of Hitchin
type. The different approaches to the double elliptic systems are
unified using the geometry behind the Mukai-Odesskii algebra.
\end{abstract}


\section{Introduction}

During the last years duality becomes a very fashionable term
denoting a lot of different phenomena mainly due to the dramatic
development in the string theory. Hence it is necessary to be very
precise when speaking on the related issues. That is why  we
formulate the subject of the paper from the very beginning - the
different dualities in the Hitchin type many-body systems as well
as their generalizations shall be discussed. These include Gaudin
system, Calogero and Toda type systems, their relativizations due
to Ruijsenaars and more general systems involving the elliptic
dependence on the momentum variables including the double elliptic
model.

The idea of some hidden duality property which could help to solve
these systems was pioneered by Ruijsenaars in \cite{ru}. It has
been actually forgotten for a decade and was revived recently
after the breakthrough in the string theory. In the recent works
\cite{fgnr,gnr} it was reformulated in more geometric terms
utilizing the fact that the phase spaces of the  corresponding
systems are the manifolds coinciding with some moduli spaces.
These moduli spaces or their close relatives enjoy a large
symmetry group due to their origin and we shall demonstrate that
part of the symmetries can be precisely formulated in terms of
duality.

Seemingly different problem concerns the separation of variables
procedure which means that the complicated interacting many-body
problems can be reduced to the collection of the identical systems
with one degree of freedom. The technical recipe for the
derivation of the separated variables in a wide range of the
dynamical systems was formulated by Sklyanin \cite{sesklyanin}.
Recently it was shown \cite{gnr} that the separation of variables
for the Hitchin type many body systems can be recognized within
the duality group. It appeared that this procedure has the T
duality transformation as the string theory counterpart.

All types of dualities can be thought of as a kind of canonical
transformation therefore the detailed structure of the phase
spaces has to be investigated to get the complete set of the
independent symmetries. In the first type of duality the initial
action-angle variables in the initial system become $p,q$
variables in the dual one. In the separation of variables
procedure one maps the initial phase space into the symmetric
power of the symplectic manifold which serves as the phase space
for the one-body problem. Finally one more duality which is
analogue of the $S$ duality in the string theory can be formulated
\cite{fgnr} as the change of the basis in the base of the
Liouville tori fibration over the space of the integrals of
motion.

The phase spaces of the integrable systems we are
interested in admit several descriptions. The
first one involves the Hamiltonian reduction
procedure for the group like
symplectic manifolds. The group approach to
the integrable many body system is known for a while \cite{olper}
and the idea to use the Hamiltonian reduction procedure
for the finite dimensional groups to derive
the Calogero type systems was invented in \cite{kks}.
The generalization along this line involves
the affine groups and algebras \cite{gn,gnru} which amounts
to the more general
Calogero and Ruijsenaars systems while the
double affine structures amount to
the elliptic Calogero and Ruijsenaars models \cite{gn2,afm}.
The part of
such systems can be considered as examples of the Hitchin systems
\cite{hitchin}.

Being reformulated in terms of the gauge theories
this approach yields the effective description of
the dynamics on the different moduli spaces. Moduli
space involved can be identified as the moduli space
of the flat $G$ connections on the torus with marked
points, moduli space of the holomorphic vector
bundles on the Riemann surfaces and moduli
spaces of instantons and monopoles. It appears
that essentially all relevant moduli spaces
can be produced from the instanton moduli spaces
moreover generically instantons have to be considered on
the noncommutative manifolds.

Another line of reasoning involves the idea of the separation of
the variables. Within this approach the integrable systems differ
by the four dimensional symplectic manifolds where the separated
variables live on. Then the dynamics is reformulated in terms of
the ADHM like description of the Hilbert scheme of points on the
surface. The simplest models like rational and trigonometric
Calogero systems can be effectively considered along this approach
however for the more complicated systems there is the obstacle due
to the lack of the ADHM like description of points on the generic
four dimensional manifolds. Nevertheless we shall formulate very
explicitly the corresponding four dimensional manifolds for all
models including double elliptic models. We shall unify two
seemingly different approaches to double elliptic two-body system
\cite{fgnr,bmmm} utilizing the Mukai-Odesskii algebra associated
with this manifold.

This paper is an extended version of the V.R. talk given  during
the Kiev NATO Advanced Research Workshop "Dynamical Symmetries of
Integrable Quantum Field Theories and Lattice models" in September
2000. We are based heavily on our papers  \cite{fgnr,gnr} but we
also present some new results which are part of our joint works
with H. Braden and A. Odesski in progress.

The work is organized as follows. In section 2 we describe the
algebraic approach based on the Hamiltonian reduction procedure
and formulate the analogue of the mirror symmetry. Some comments
concerning $S$ duality transformation and the quantum case are
presented. In section 3 we explain the geometry behind the
separation of the variables procedure and clarify the
corresponding four dimensional manifolds for all systems under
consideration. Some new results concerning the double elliptic
system and its possible generalizations are presented. In Section
4 we conclude with some remarks concerning the application of
these dualities to the physical models.

\section{"Mirror" symmetry  in dynamical systems}

The first type of duality we would like to discuss concerns
the dualities between the pair of dynamical systems \cite{ru,fgnr}.
It has been formulated by Ruijsenaars \cite{ru} in the
context of the transition to the action-angle variables.
In \cite{fgnr} the general procedure for the analogous
symmetry within integrable system
in terms of Hamiltonian and Poissonian
reductions was formulated. Symmetry
maps one  dynamical system with coordinates $x_i$    to
another one whose coordinates coincide with the action
variables of the initial system and vice versa. It is
essential that transition from the Hamiltonians in the
initial dynamical system to the ones in the dual system
can be also formulated in geometrical terms.

Qualitatively this symmetry is even more
transparent in terms
of separated variables.
The proper object to discuss in the separated variables
is the hyperkahler
four dimensional manifold
which provides  the phase space for one degree of freedom. In the most
general situation under consideration the manifold involves two tori  or
elliptically fibered K3 manifold.
One torus provides momenta while
the second yields the coordinates. The duality at hands
actually interchanges momentum and coordinate tori
and in some case self-duality is expected.
Since the manifold typically has the structure
of the bundle the interchange of the base and the fiber
is highly nontrivial operation.
All other cases correspond to some degeneration.

We will discuss mainly classical
case with only few comments on the quantum picture. Since
the wave functions in the Hitchin like systems can be
identified with some solutions to the KZ or qKZ equations
the quantum duality would mean some relation between
solutions to the rational, trigonometric or elliptic qKZ
equations. Recently the proper symmetries for KZ
and qKZ equations
where found in \cite{qdual}. To find the proper symmetry for the
elliptic case still remains a challenging problem.

\subsection{Two-body system ($SU(2)$)}
Let us discuss first two-body system corresponding to $SU(2)$ case.
Two-particle systems which we are going to consider  reduce
to a one-dimensional problem.
The action-angle variables can  be written explicitly  and
the
dual system emerges immediately once the natural Hamiltonians
are chosen. The problem is the following. Suppose
the phase space is coordinatized by $(p,q)$. The dual Hamiltonian
is a function of $q$
expressed in terms of $I, \varphi$, where $I, \varphi$ are the action-angle
variables of the original system :  $H_{D}(I, \varphi) = H_{D}(q)$.

Consider as example elliptic Calogero model which is rather illustrative
since already in this case the duality provides an important tool
to generate new integrable systems with the elliptic dependence on momenta.
The Hamiltonian reads:
\beq
H(p,q) = {{p^{2}}\over{2}}  + {\nu^{2}} {\wwp}.
\eeq
Here $p,q$ are complex, $\wwp$ is the Weierstrass
function on the elliptic curve $E_{\tau}$:
\beq
{\wwp} = {1\over{q^{2}}} + \sum_{\matrix{&(m,n) \in
{Z^{2}} \cr  &(m,n) \neq  (0,0) \cr}}
{1\over{(q + m\pi  +n \tau\pi )^{2}}} - {1\over{(m\pi  +n \tau\pi)^{
2}}}
\eeq
Let us introduce the Weierstrass notations:
$x = {\wwp}$, $y = {\wwp}^{\prime}$.
We have an equation defining the curve $E_{\tau}$:
\beq
y^{2} = 4x^{3} - g_{2}(\tau) x -  g_{3}(\tau) =
4 \prod_{i=1}^{3} (x - e_{i}), \quad \sum_{i=1}^{3} e_{i} = 0
\eeq
The holomorphic differential $dq$ on $E_{\tau}$ equals $dq = dx/y$.
Introduce the variable $e_{0} = 2E/{\nu}^{2}$.
The action variable is one of the periods of the
differential ${pdq}\over{2\pi}$ on the curve
$E = H(p,q)$ :
\beq
I = {1\over{2\pi}} \oint_{A} \sqrt{2(E - {\nu}^{2} {\wwp}) } =
{1\over{4\pi i}} \oint_{A}
{{dx \sqrt{x - e_{0}}}\over{\sqrt{(x-e_{1})(x-e_{2})(x-e_{3})}}}
\eeq
The angle variable can be determined from the condition
$dp \wedge dq = dI \wedge d\varphi$:
\beq
d\varphi = {1\over{2i T(E)}} {{dx}\over
{\sqrt{\prod_{i=0}^{3}(x - e_{i})}}}
\eeq
where $T(E)$ normalizes $d\varphi$ in such a way that
the  $A$ period of $d\varphi$ is
equal to $2\pi$:
\beq
T(E) ={1\over{4\pi i }}
\oint_{A} {{dx}\over{\sqrt{\prod_{i=0}^{3} (x - e_{i})}}}
\eeq
Thus:
\beq
2i T(E)
d\varphi =  {{dx}\over{\sqrt{4 \prod_{i=0}^{3}
(x - e_{i})}}}
\eeq
\beq
\omega d\varphi =
{{dt}\over{\sqrt{4 \prod_{i=1}^{3} ( t- t_{i})}}}
\eeq
where
\beq
\omega = - 2i T(E) \sqrt{e_{01}e_{02}e_{03}}   =
{1\over{2\pi}} \oint_{A}  {{dt}\over{\sqrt{4 \prod_{i=1}^{3}
( t- t_{i})}}}
\eeq

\beq
t=  {1\over{x-e_{0}}} + {1\over{3}}
\sum_{i=1}^{3} {1\over{e_{0i}}} ; t_{i} = {1\over{3}}
\sum_{j=1}^{3} {{e_{ji}}\over{e_{0i} e_{0j}}}
\eeq
where $e_{ij} =  e_{i} - e_{j}$

Introduce a meromorphic function on $E_{\tau}$:
\beq
{\widehat {cn}}_{\tau}(z) = \sqrt{{x-e_{1}}\over{x-e_{3}}}
\eeq
where $z$ has periods $2\pi$ and $2\pi \tau$.

Then we have:
\beq
H_{D}(I, \varphi) = {\widehat {cn}}_{\tau}(z) =
{\widehat {cn}}_{\tau_{E}}({\varphi})
\sqrt{1 -{{{\nu}^{2}e_{13}}\over{2E - {\nu}^{2}e_{3}}}}
\eeq
where $\tau_{E}$ is the modular parameter of the relevant spectral curve
$v^{2} = 4 \prod_{i=1}^{3} ( t - t_{i} )$:
\beq
\tau_{E} = \Bigl( \oint_{B} {{dt}\over{\sqrt{4 \prod_{i=1}^{3}
(t - t_{i})}}}\Bigr) \Bigr/ \Bigl( \oint_{A}
{{dt}\over{\sqrt{4 \prod_{i=1}^{3}
(t - t_{i})}}} \Bigr).
\eeq
For large $I$, $2E(I) \sim I^{2}$.

Therefore  the elliptic Calogero
model with rational dependence on momentum and elliptic on
coordinate   maps into the "mirror" dual system with
elliptic dependence on momentum and rational on coordinate.
The generalization to the elliptic Ruijsenaars
model is quite straightforward \cite{fgnr,bmmm} and
the system dual to it manifests the elliptic dependence
on the momentum and trigonometric on coordinate.

The most general system involves elliptic dependence both on
coordinate and momenta. The Hamiltonian of the double elliptic
system in the form \beq H(p,q)=\alpha(q|k)cn(p
\beta(q|k,\tilde{k})| \frac{\tilde{k} \alpha(q|k)}{ \beta(q|k) })
\eeq where $\alpha(q|k)=\sqrt{1- \frac{g^2}{sn^2(q|k)}}$ and
$\beta(q|k,\tilde{k})=\sqrt{1- \frac{g^2\tilde{k}}{sn^2(q|k)}}$,
has been found using the self -duality argument in \cite{bmmm}.

It is instructive to compare the Hamiltonian above with the double
elliptic system suggested in \cite{fgnr}. The four dimensional
manifold which provides the phase space in that paper was
identified with the elliptically fibered K3 manifold. The
suggested Hamiltonian was linear in the coordinate on the base of
the fibration. The Hamiltonian of the dual system was linear in
the coordinate on the fiber. We shall see later that being
considered in the separated variables double elliptic system
involves the Mukai-Odesskii algebra on the intersection of
quadrics. With the natural Poisson bracket algebra on this
manifold the Hamiltonian written in the Darboux variables above
exactly coincides with the coordinate on the base of the elliptic
fibration while the dual Hamiltonian is linear in the coordinate
of the fiber. Therefore two forms of the double elliptic system
are essentially equivalent.

 This approach dictates that the
manifold in \cite{bmmm} is the double covering of the elliptically
fibered rational surface. Moreover in these terms the Hamiltonian
of the direct system is the coordinate on the fiber while the one
for the dual system with the coordinate on the base. Therefore we
could consider the same arguments for the double elliptic system
from \cite{fgnr} in the case of the rational elliptic fibrations,
(incomplete) hyperelliptic curves of genus $2$ (a typical fiber of
the dual system) and the double cover looks like a hyperelliptic
Jacobian of the genus 2 curve (this gives us a direct imbedding of
our picture in the Beauville - Mukai systems a la K. Takasaki
\cite{taka}). We can illustrate the situation with the following
diagramme of double elliptic system  with the interchange of the
direct and dual systems.

$$
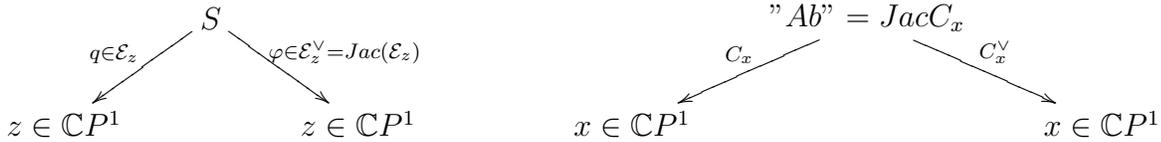
\xymatrix{
&S\ar[dl]_{q\in \mathcal{E}_z} \ar[dr]^{\varphi\in
  \mathcal{E}_{z}^{\vee} = Jac(\mathcal{E}_z)}&  & & &"Ab"=Jac C_x\ar[dl]_{C_x}\ar[dr]^{C_{x}^\vee}&\\
z\in \mathbb{C}P^1&&z\in \mathbb{C}P^1&&x\in \mathbb{C}P^1&&x\in\mathbb{C}P^1\\
}
$$

Moreover , now we are able to show the coincidence of the "naive"
definition of the duality originated in the "duality relation" of
\cite{bmmm} arising from the anticanonicity condition:
$$
dP\wedge dQ = - dp\wedge dq.
$$

We consider two "dual" elliptic fibrations associated with the
double elliptic system of
\cite{bmmm}
system and with its dual as projective Weierstrass cubic families in
$\mathbb{C}P^2$:
$$
S\subset\mathbb{C}P^2 : y^2 = z^3 + f(x)z + g(x), x\in\mathbb{C}P^1,
$$
and $z,x$ are nonhomogeneous coordinates in $\mathbb{C}P^2$, with the canonical
symplectic form $\Omega = {{dz\wedge dx}\over{y}}$
and
$$
{\tilde S}\subset\mathbb{C}P^2 : y^2 = x^3 + {\tilde f}(z)x +{\tilde g}(z), z\in\mathbb{C}P^1,
$$
and $x,z$ are nonhomogeneous coordinates in $\mathbb{C}P^2$, with the canonical
symplectic form $\Omega = {{dx\wedge dz}\over{y}}$.

Now the local action coordinate $I(x)$ is computed (as we had argued in
\cite{fgnr}) as
$$
dI(x) = T(x)dx = \left(1/2\pi \oint_{A_x}dz/y\right)dx,
$$
where $[A_x]\in H_1({\mathcal E}_x, {\mathbb Z})$ is a chosen
$A$-cycle. Analogously, the dual action coordinate $I^{D}(z)$ is
satisfied to
$$
dI^{D}(z) = T^{D}(z)dz = \left(1/2\pi \oint_{L_z}dx/y\right)dz,
$$
where $[L_z]\in H_1({\mathcal C}_z, {\mathbb Z})$.

Now we have the following chain of transformations:
$$
\Omega ={{dz\wedge dx}\over{y}}= - {{dx\wedge dz}\over{y}},
$$
$$
{{dz}\over{y}}\wedge {{dI}\over{T(x)}} = - {{dx}\over{y}}\wedge {{dI^{D}}\over{T^{D}(z)}},
$$
or
$$
{{dz}\over{yT(x) }}\wedge dI = - {{dx}\over{yT^{D}(z) }}\wedge dI^{D}.
$$
Now (using the expressions for the angle variables $\varphi ={{dz}\over{yT(x) }}$ and
${\varphi}^D = {{dx}\over{yT^{D}(z) }}$) we obtain
$$
d\varphi \wedge dI = - d{\varphi}^D\wedge dI^{D}.
$$

Keeping in mind the interpretation of \cite{bmmm}, we have
from the equality above
$$
K(k) dp^{Jac}\wedge dI = - K(\tilde k) dp^{{Jac}^{\vee}}\wedge dI^{D},
$$
or
$$
dP\wedge dQ = - dp\wedge dq
$$
as it was supposed in \cite{bmmm}!

\subsection{Many-body systems}
Now we would like to demonstrate how the "mirror" transform can be
formulated in terms of Hamiltonian or Poissonian reduction
procedure. It appears that it corresponds in some sense to the
simultaneous change of the gauge fixing and Hamiltonians. More
clear meaning of these words will be clear from the examples
below.

We summarize the systems
and their duals in rational
and trigonometric cases in the following table:
\beq
\matrix{& &{\rm rat. CM} &
\leftrightarrow & {\rm rat. CM} & &\cr
& R \to 0 &   \uparrow  &
                 &  \uparrow       & \beta \to 0& \cr
& &{\rm trig. CM} & \leftrightarrow & {\rm rat. RS} & &\cr
& \beta \to 0 &   \uparrow  &
                        &  \uparrow       &R \to 0&\cr
& &{\rm trig. RS} & \leftrightarrow & {\rm tri
g. RS} & &\cr}
\eeq

Here $CM$ denotes ${\it Calogero-Moser}$ models  and $RS$ stands
for ${\it Ruijsenaars-Schneider}$. The parameters $R$ and $\beta$
here are the radius of the circle the coordinates of the particles
take values in and the inverse speed of light respectively. The
horizontal arrows in this table are the dualities, relating the
systems on the both sides. We notice that the duality
transformations form a group which  in the case of self-dual
systems listed here contains ${\rm SL}_{2}(Z)$. The generator $S$
is the horizontal arrow described below, while the $T$ generator
is in fact a certain finite time evolution of the original system
(which is always a symplectomorphism, which maps the integrable
system to the dual one). \noindent Throughout this section
$q_{ij}$ denotes $q_{i} - q_{j}$.

Let us take trigonometric Ruijsenaars model as an example
since it can be effectively described via Poissonian
and Hamiltonian reductions. Let us start with the Poissonian one.
Consider the space
${\CA}_{{\bf T}^{2}}$ of
$SU(N)$ gauge fields $A$ on a two-torus
${\bf T}^{2} = {\bf S}^{1} \times {\bf S}^{1}$.
Let the circumferences of the circles
be $R$
and $\beta$.
The space  ${\CA}_{{\bf T}^{2}}$ is acted on by a gauge group $\CG$ ,
which preserves a symplectic form
\beq
\Omega = {k\over{4{\pi}^{2}}}
\int {\Tr} \delta A \wedge \delta A,
\eeq
with $k$ being an arbitrary real
number for now.
The gauge group acts via evaluation at
some point $p \in {\bf T}^{2}$
on any coadjoint orbit $\CO$ of $G$, in particular, on $\CO =
{\IC\IP}^{N-1}$.
Let $(e_{1} :  \dots : e_{N})$
be the homogeneous
coordinates on $\CO$. Then the moment map for the action of $\CG$
on $\CA_{{\bf T}^{2}} \times \CO$ is
\beq
k F_{A} + J \delta^{2}(p),  \quad J_{ij}
= {\sqrt -1}{\nu} (  \delta_{ij} - e_{i} e_{j}^{*})
\eeq
$F_{A}$ being the curvature
two-form. Here we think of $e_{i}$ as
being the coordinates on $\IC^{N}$
constrained so that $\sum_{i} \vert e_{i} \vert^2  =N$ and
considered up to the multiplication
by a common phase factor.

Let us
provide a certain amount of commuting Hamiltonians. Obviously,
the
eigen-values of the monodromy of $A$
along any fixed loop on ${\bf T}^{2}$
commute
with themselves. We consider the
reduction at the zero level of the moment
map.  We have at least $N-1$ functionally independent commuting
functions on
the reduced phase space $\CM_{\nu}$.

Let us estimate the dimension
of $\CM_{\nu}$.
If $\nu= 0$ then the moment equation forces the connection to be
flat and therefore
its gauge orbits are parameterized by the conjugacy
classes of the monodromies
around two non-contractible cycles on ${\bf T}^{2}$: $A$
and $B$. Since the fundamental
group $\pi_{1} ({\bf T}^{2})$ of ${\bf T}^{2}$
is abelian $A$ and $B$ are to
commute. Hence they are simultaneously diagonalizable, which makes
${\CM}_{0}$ a
$2(N-1)$ dimensional manifold. Notice that the generic point on the
quotient
space has a nontrivial stabilizer, isomorphic to the maximal torus
$T
$ of $SU(N)$. Now, in the presence of $\CO$
the moment equation implies
that the connection $A$ is flat outside of $p$ and has a
nontrivial
monodromy around $p$. Thus:
\beq
ABA^{-1}B^{-1} = {\exp}(R\beta J)
\eeq
(the factor $R\beta$ comes from the normalization of the
delta-function ).
If we diagonalize
$A$, then $B$ is uniquely
reconstructed up to the right
multiplication by the elements of $T$. The
potential degrees of freedom in $J$ are
"eaten" up by the former stabilizer
$T$ of a flat connection: if we conjugate both
$A$ and
$B$ by an element $t
\in T$ then $J$ gets conjugated. Now, it is important that
$\CO$ has
dimension $2(N-1)$. The reduction of
$\CO$ with respect to $T$ consists of a
point and does not contribute to the
dimension of $\CM_{\nu}$. Thereby we
expect to get an integrable system.
Without doing any computations we already
know
that we get a pair of dual systems. Indeed, we may choose as the set of
coordinates the eigen-values
of $A$ or the eigen-values of $B$.

The two-dimensional
picture has the advantage that the geometry of the problem
suggest the
$SL_{2}(Z)$-like duality. Consider the operations $S$ and $T$  realized
as:
\beq
S: (A, B) \mapsto (ABA^{-1}, A^{-1}); \quad T: (A,B)
\mapsto (A,  BA)
\eeq
which correspond to the freedom of  choice  of generators  in
the fundamental group of a two-torus. Notice that
both $S$ and $T$ preserve the commutator $ABA^{-1} B^{-1}$ and
commute with the action of the gauge group.
The group $\Gamma$ generated by $S$ and $T$
in the limit $\beta, R \to 0 $
contracts to ${\rm SL}_{2}(Z)$ in a sense that
we get the
transformations
by expanding
$$
A = 1 + \beta P + \ldots, \quad B = 1 + R Q +
\ldots
$$
for $R, \beta \to 0$.

To perform the Hamiltonian reduction  replace the space of
two dimensional gauge fields by the
cotangent space to the (central extension of)
loop group:
$$
T^{*}{\hat G}   = \{ ( g(x), k\p_{x} + P(x) ) \}
$$
The relation to the two dimensional
construction
is the following. Choose a non-contractible circle $\bf S^{1}$
on the two-torus
which does not pass through the marked point $p$. Let $x,y$
be the coordinates on the torus
and $y=0$ is the equation of the $\bf S^{1}$.
The periodicity  of $x$ is $\beta$ and that of $y$ is $R$.
Then
$$
P(x) = A_{x}(x,0),
g(x) =P\exp\int_{0}^{R} A_{y}(x,y) dy.
$$
The moment map equation looks as follows:
\beq
k g^{-1} \p_{x}
g + g^{-1}Pg - P = J \delta(x),
\eeq
with $k = {1\over{R\beta}}$. The solution of
this equation in the gauge $P = {\rm diag}(q_{1}, \ldots, q_{N})$
leads to
the Lax operator $A = g(0)$ with $R,\beta$
exchanged. On the other hand,
if we  diagonalize
$g(x)$:
\beq
g(x) =  {\rm diag} \left( z_{1} =
e^{{\sqrt -1}R q_{1}}, \ldots, z_{N} = e^{{\sqrt -1}R q_{N}} \right)
\eeq
then a similar calculation
leads to the Lax operator
$$
B = P\exp\oint{1\over{k}} P(x)dx =  {\rm diag} (
e^{{\sqrt -1}\theta_{i} } ) \exp {\sqrt -1}R\beta\nu {\rm r}
$$
with
$$
{\rm r}_{ij} =
{1\over{1- e^{{\sqrt -1}Rq_{ji}}}}, i\neq j; \quad {\rm r}_{ii} = - \sum_{j\neq i}
{\rm r}_{ij}
$$
thereby establishing the duality $A \leftrightarrow B$
explicitly.

Let us briefly rephrase the discussion above in a more physical
language.
When Yang-Mills theory is formulated on a cylinder
with the insertion of an appropriate time-like Wilson line, it
is equivalent to the Sutherland model describing a collection
of $N$ particles on a circle. The observables ${\Tr} \phi^k$
are precisely the integrals of motion of
this system.
One can look at other supercharges as well. In particular,
when the theory is formulated on a cylinder there is another
class of observables annihilated
by a supercharge. One can arrange the
combination
of supercharges which  will annihilate the Wilson loop operator. By repeating
the procedure similar to the one in  one arrives at the
quantum mechanical theory whose Hamiltonians are generated
by the spatial Wilson loops. This model is nothing
but the rational Ruijsenaars-Schneider many-body system.

The self-duality of trigonometric Ruijsenaars system has even
more transparent physical meaning. Namely, the field theory
whose quantum mechanical avatar is the Ruijsenaars system is
three dimensional Chern-Simons theory on
${\bf T}^{2} \times {\bf R}^{1}$ with the insertion of an appropriate
temporal Wilson line and spatial Wilson loop. It is the freedom
to place the latter which leads to several equivalent theories.
The group of (self-)dualities of this model is very big
and is generated by the transformations $S$ and $T$ .

Let us remark that the choice of the dual Hamiltonians
remains a delicate issue. At a moment we can mention
two approaches to this  problem. One of them
exploits the
embedding of the integrable many-body systems in
the Toda lattice equations as the dynamics of the
zeros of the corresponding $\tau$ functions \cite{mm}.
In the second approach one defines the
Toda lattice system in terms of the deformations  of the
Lagrangian manifolds of the generic Hamiltonian systems
\cite{rg}. In such approach the dual Hamiltonians
have a simple structure in terms of the
creation operators defined in the initial system \cite{rg}.
Certainly this problem  deserves further
investigation.

\subsection{Quantum systems.}

Our discussion so far concerned the classical systems. However
duality is expected to be a powerful tool to deal with the quantum
problems. This point has been recognized by Ruijsenaars in his
early papers. The very idea is that the Hamiltonian and its dual
have the common wave function and this fact could be exploited to
consider one or another eigenvalue equations with respect to
different variables. Since the wave functions in the systems under
consideration admits the simple group like realization one could
expect that the quantum duality can be formulated purely in the
group theory terms too. Recent activity in this direction confirms
such expectations \cite{qdual}.

Here we work out a few  examples of
quantum dual systems to illustrate the general picture.
 The Hamiltonian of the oscillator
 quantizes to:
\beq
{\hat H} = {-1/2}{{{\p}^{2}}\over{{\p} q^{2}}} +
{{{\omega}^{2} q^{2}}\over{2}}
\eeq
Its
normalized eigen-functions are  :
\beq
{\hat H}
{\psi}_{n}  =  {\omega} (n + {1/2}) \psi_{n}
\eeq
\beq
{\psi}_{n} (q) =  \bigl(
{{\omega}\over{\pi}} \bigr)^{1/4}  {{
e^{-{{\omega
q^{2}}\over{2}}}
}\over{2^{n/2}\sqrt{n!}}} H_{n}(q \sqrt{\omega})
\eeq
where
$H_{n}(\xi)$ is the Hermite polynomial:
$H_{n}(\xi) = e^{\xi^{2}} (-
\p_{\xi})^{n} e^{-{\xi}^{2}}$.
Using this representation of the
wave-function one can easily obtain a reccurence
relation :
\beq
\sqrt{n+1} {\psi}_{n+1}(q) + \sqrt{n} {\psi}_{n-1}
(x) = \sqrt{2\omega} q {\psi}_{n}(x)
\eeq
It means that $\psi_{n}(q)$ is an
eigen-function of the following
difference operator:
\beq
{\hat H}_{D}
= T_{+} \sqrt{n} + \sqrt{n} T_{-}, \quad T_{\pm} = e^{\pm {{\p}\over{\p
n}}}
\eeq
acting on the  subscript $n$. It is easy to recognize   the
quantized version of the dual system.

Another example  involves the Hamiltonian:
\beq
{\hat H} = {-\half }{{{\p}^{2}}\over{{\p}
q^{2}}} + {{{\nu}({\nu} -1)}\over{2 \sin^{2}(q)}}
\eeq
Its normalized
eigen-functions are :
\beq
{\hat H} \psi_{n}  =
{{n^{2}}\over{2}} \psi_{n}
\eeq
\beq
\psi_{n}(q) = \sin^{\nu}(q)   \sqrt{n
{{(n-{\nu})!}\over{(n+{\nu}-1)!}}} \Pi_{n-\half}^{\nu -
\half}\bigl(\cos(q)\bigr)
\eeq
\beq
\Pi_{l}^{m}(x) = {1\over{l!}} {\p}_{x}^{l+m}
\bigl({{x^{2}-1}\over{2}}\bigr)^{l}
\eeq
For simplicity we take $\nu$ and
$n$ to be half-integers.
One can change $\nu \to - \nu - 1$ to get
another
eigen-function with the same eigenvalue.
Using the fact that the
generating function for $\Pi_{l}^{0}$'s is
\beq
Z(y,x) =
\sum_{l=0}^{\infty} y^{l} \Pi_{l}^{0} = {1\over{\sqrt{1 - 2xy + y^2}}}
\eeq
one derives the
recurrence relations
using two obvious equations:
\beq
{(x-y)\p_{x} Z = y\p_{y} Z}
\eeq
\beq
{(1 - 2xy + y^{2})\p_{y} Z = (x-y)Z}
\eeq
Next it implies:
\beq
(y\p_{y} - m) \p_{x}^{m} Z = (x- y) \p^{m+1}_{x} Z
\eeq
and hence yields:
\beq
{ \bigl( (1 - 2xy + y^{2}) \p_{y}  + y - x \bigr) \p^{m}_{x}Z
= m ( 1 + 2y\p_{y}) \p^{m-1}_{x} Z}
\eeq

Combination of those two gives rise the desired relation.

\beq
x \Pi_{l}^{m} = {{l+1-m}\over{2l+1}} \Pi_{l+1}^{m} + {{l+m}\over{2l+1}}
\Pi_{l-1}^{m}
\eeq
\beq
\cos(q) \psi_{n} = {\half} \Bigl( \sqrt{1-
{{\nu}({\nu}-1)}\over{n(n+1)}}\psi_{n+1}
+\sqrt{1-
{{\nu}({\nu}-1)}\over{n(n-1)}}\psi_{n-1} \Bigr)
\eeq
that is $\psi_{n}$ is an
eigen-function of the finite-difference operator
acting on the $n$
subscript:
\beq
{\hat H}_{D} \psi(q) = \cos(q) \psi(q)
\eeq
\beq
{\hat H}_{D} = T_{+} \sqrt{1- {{{\nu}({\nu}-1)}\over{n(n-1)}} }+
\sqrt{1- {{{\nu}({\nu}-1)}\over{n(n-1)}} }T_{-}
\eeq
which is  a quantum
version of rational Ruijsenaars model.

Summarizing, when the system with trigonometric dependence on
momentum is quantized its Hamiltonian becomes a finite difference
operator. The wave functions become the functions of the discrete
variables. The origin of this is in the Bohr-Sommerfeld
quantization condition. Indeed, since the trigonometric dependence
of momenta implies that the leaves of the polarization are compact
and moreover non-simply connected the covariantly constant
sections of the prequantization connection along the polarization
fiber generically ceases to exist. It is only for special
``quantized'' values of the action variables that the section
exists. In the elliptic case the quantum dual Hamiltonian is going
to be a difference operator of the infinite order.

\section{S-duality}

Now let us explain that S-duality
well established in the field theory  also
has clear counterpart
in the holomorphic dynamical system.

The action variables in dynamical system
are the integrals of meromorphic differential $\lambda$ over
the $A$-cycles on the spectral curve.
The reason for the $B$-cycles to be
discarded is simply the fact that the $B$-periods of $\lambda$
are not independent of the $A$-periods. On the other
hand, one can choose as the independent periods the
integrals of $\lambda$ over any lagrangian
subspace in $H_{1}({\bf T}_{b}; Z)$.

This leads to the following structure of the action variables in
the holomorphic setting. Locally over a disc in $B$ one chooses a
basis in $H_{1}$ of the fiber together with the set of $A$-cycles.
This choice may differ over another disc. Over the intersection of
these discs one has a $Sp(2m, \mathbb {Z})$ transformation
relating the bases. Altogether they form an $Sp(2m, \mathbb {Z})$
bundle. It is an easy exercise on the properties of the period
matrix that the two form: \beq dI^{i} \wedge dI^{D}_{i} \eeq
vanishes. Therefore one can always locally  find a function $\CF$
- {\it prepotential}, such that: \beq I^{D}_{i} = {{\p
\CF}\over{\p I^{i}}} \eeq The angle variables are uniquely
reconstructed once the action variables are known.

To illustrate the meaning of the action-action
duality we look at the
two-body system, relevant for the
$SU(2)$ $\CN=2$ supersymmetric
gauge theory :
\beq
H =
{{p^{2}}\over{2}} + \Lambda^{2} \cos (q)
\eeq
with $\Lambda^{2}$ being a complex
number - the coupling constant of a two-body
problem and at the same time a
dynamically generated scale of the gauge theory.
The action variable is
given by one of the periods of the differential $pdq$.
Let us introduce more
notations:
$x = \cos (q)$, $y = {{p\sin (q)}\over{\sqrt{-2}\Lambda}}$, $u =
{{H}\over{\Lambda^{2}}}$. Then the
spectral curve, associated to the system
which is also a level set of the Hamiltonian
can be written as
follows:
\beq
y^{2} = (x - u) (x^{2}-1)
\eeq
which is exactly
Seiberg-Witten curve .
The periods are:
\beq
I = \int_{-1}^{1} \sqrt{{x-u}\over{x^{2}-1}} {dx},
I^{D} = \int_{1}^{u} \sqrt{{x-u}\over{x^{2}-1}} {dx}
\eeq
They obey  Picard-Fuchs equation:
$$
\left( {{d^{2}}\over{du^{2}}} +
 {1\over{4(u^{2}-1)}} \right) \pmatrix{& I\cr& I^{D}\cr} = 0
$$
which can be
used to write down an asymptotic expansion of the action
variable near
$u=\infty$ or $u= \pm 1$ as well as that of prepotential . The
duality is manifested in the fact that
near $u =\infty$ (which
corresponds to the high energy scattering in the two-body problem
) the appropriate action
variable
is $I$ (it experiences a monodromy $I \to - I$ as $u$ goes around
$\infty$), while
near $u = 1$ (which corresponds to the dynamics of the
two-body
system near the top of the potential )
the appropriate variable is $I^{D}$ (which  is actually well
defined near $u=1$ point) . The monodromy
invariant combination of the
periods \cite{matone}:
\beq
II^{D} - 2\CF = u
\eeq
can be chosen
as a global coordinate on the space of integrals of motion .
At $u \to \infty$ the prepotential has an expansion of the form:
$$
\CF \sim {\half}  u \log u + \ldots \sim I^{2}
\log I + \sum_{n}{{f_{n}}\over{n}} I^{2-4n}
$$

Let us emphasize that S-duality maps the dynamical
system to itself. We have seen that the notion of prepotential can be
introduces for any holomorphic many-body system
however its physical meaning  as well as its properties
deserve further investigation.

\section{$T$ duality and separation of variables}
Let us consider  the analogue of $T$ duality in the Hitchin like
systems \cite{gnr}. It appears that the proper analogue of $T$
duality can be identified with the separation of variables in the
dynamical systems. A way of  solving a problem with many degrees
of freedom is to reduce it to the problem with the smaller number
of degrees of freedom. The solvable models  allow to reduce the
original system with $N$ degrees of freedom to $N$ systems with
$1$ degree of freedom which reduce to quadratures. This approach
is called a separation of variables ( SoV). Recently, E.~ Sklyanin
formulated ``magic recipe'' for the  SoV in the large class of
quantum integrable models with a Lax representation
\cite{sesklyanin}. The method reduces in the classical case to the
technique of separation of variables using poles of the
Baker-Akhiezer function (see also \cite{kuz,kripho}) for recent
developments and more references). The basic strategy of this
method is to look at the Lax   eigen-vector ( which is the
Baker-Akhiezer function) $\Psi (z, \lambda)$: \beq L(z) \Psi(z,
\lambda) = \lambda (z) \Psi(z, \lambda) \eeq with some choice of
normalization. The poles $z_{i}$ of $\Psi(z, \lambda)$ together
with the eigenvalues $\lambda_{i} = \lambda(z_{i})$ are the
separated variables. In all the examples studied so far the most
naive way of normalization leads to the canonically conjugate
coordinates $\lambda_{i}, z_{i}$.

Remind that the phase space for the  Hitchin system can be
identified with the cotangent bundle to the moduli space of
holomorphic vector  bundle $T^{*}M$ on the surface $\Sigma$. The
following symplectomorphisms can be identified with the separation
of variables procedure. The phase space above allows two more
formulations; as the pair $(C,\cal{L})$ where C is the spectral
curve of the dynamical system and $\cal{L}$ is the linear bundle
or  as the Hilbert scheme of points on $T^{*}\Sigma $ where the
number of points follows from the rank of the gauge group. It is
the last formulation provides the separated variables. The role of
Hilbert schemes on   $T^{*}\Sigma $   in context of Hitchin system
was established for the surfaces without marked points in
\cite{hurt} and generalized for the systems of Calogero types in
\cite{wilson, nakajima, gnr}.

\subsection{Gaudin model}
Let us present the explicit realization of the separation of
variables procedure in terms of BA function in the Gaudin model.
Consider the space \beq \CM = \left( \CO_{1} \times \ldots \times
\CO_{k} \right) // G \eeq where $\CO_{l}$ are the complex
coadjoint orbits of $G = {\rm SL}_{N}({\IC})$ and the symplectic
quotient is taken with respect to the diagonal action of $G$.

This moduli space parameterizes Higgs pairs on ${\IP}^{1}$ with
singularities at the marked points $z_{i} \in {\IP}^1, \quad i =1,
\ldots, k$. This is a natural analogue of the Hitchin space for
genus zero. The connection to the bundles on $\IP^{1}$ comes about
as follows: consider the moduli space of Higgs pairs: $(\pb_{A},
\phi)$ where $\phi$ is a {\it meromorphic} section of ${\rm ad}
(V) \otimes \CO(-2)$, with the restriction that ${\res}_{z =
z_{i}} \phi \in \CO_{i}$. The moduli space is isomorphic to
$\CM$.This space is integrable system and the separation of
variables in it has been studied in \cite{sesklyanin,frenkel,efr}.
Indeed, consider the solution to the equation

\beq
\pb_{A}\phi = \sum_{i} \mu_{i}^{c} \delta^{(2)}(z - z_{i})
\eeq
in the gauge where $\bar A=0$ which exists due to the Grothendieck's
theorem on stable holomorphic bundles on $\bf CP^1$ .
We get:
\beq
\phi(z) = \sum_{i} {{\mu_{i}^{c}}\over{z - z_{i}}}
\eeq
provided that $\sum_{i} \mu_{i}^{c} =0$
and is defined up to a global conjugation by an element
of $G$ hence the Hamiltonian reduction in $\CM$.
Now, consider the following polynomial:
\beq
{\Det}\left(\lambda - \phi(z)\right) = \sum_{i,l} A_{i,l}
\lambda^{i} z^{-l}
\eeq
The number of functionally
independent coefficients $A_{i,l}$ is precisely
equal to
$$
k \left( {{N (N-1)}\over{2}} \right) + 1 - N^{2}
$$

In the case $N=2$ the coadjoint
orbits
$\CO_{i}$ can be explicitly described as the surfaces in $\IC^{3}$ given
by the equations:
\beq
\CO_{i} : Z_{i}^{2}+ X_{i}^{+}X_{i}^{-} = \zeta_{i}^{2}
\eeq
with the symplectic forms:
\beq
\omega_{i} = {{dZ_{i} \wedge dX_{i}^{+}}\over{X_{i}^{-}}}
\eeq
and the complex moment maps:

$$
\mu_{i}^{c} =
\left(
\begin{array}{cc}
Z_{i} &   X_{i}^{+}\\
X_{i}^{-}  & - Z_{i}
\end{array}
\right)
$$

The phase space of our interest is $\CP = \times_{i=1}^{k} \CO_{i} //
SL_{2}$.
It is convenient to work with a somewhat larger space $\CP_{0} =
\times_{i=1}^{k}\CO_{i} / \IC^{*}$,
where $\IC^{*} \in SL_{2}(\IC)$ acts as follows:
$$
t : \left( Z_{i}, X_{i}^{+}, X_{i}^{-} \right) \mapsto
\left( Z_{i}, t X_{i}^{+} , t^{-1} X_{i}^{-} \right)
$$

The moment map of the torus $\IC^{*}$ action is simply $\sum_{i} Z_{i}$.
The complex dimension of $\CP_{0}$ is equal to $2(k-1)$.
The Hamiltonians are obtained by expanding the
quadratic invariant:

\beq
T(z) = {\half} {\Tr} \phi(z)^{2}, \qquad
\phi(z)  = \sum_{i}
{{\mu_{i}^{c}}\over{z-z_{i}}}
\eeq

\be
 T(z) = \sum_{i}  {{\zeta_{i}^{2}}\over{(z-z_{i})^{2}}} +
\sum_{i} {{H_{i}}\over{z-z_{i}}} \\
 H_{i} = {\half} \sum_{j \neq i} {{X_{i}^{+}X_{j}^{-} + X_{i}^{-}X_{j}^{+}
+ 2 Z_{i} Z_{j} }\over{z_{i} -
z_{j}}}
\ee
The separation of variables proceeds in this case as follows:
write $\phi(z)$ as
$$
\phi(z) =
\left(
\begin{array}{cc}
h(z) & f(z)\\
e(z) & - h(z)
\end{array}
\right).
$$
Then Baker-Akhiezer function is given explicitly by:
$$
\Psi(z) =
\left(
\begin{array}{c}
\psi_{+} \\
\psi_{-}
\end{array}
\right),
\quad \psi_{+} = f, \quad \psi_{-} = \sqrt{h^{2} + ef} - h
$$
and its zeroes are the roots of the equation
\beq
f(p_{l}) = 0 \Leftrightarrow  \sum_{i}
{{ X_{i}^{+}}\over{p_{l} - z_{i}}} = 0, \quad l = 1, \ldots, k-1
\eeq
The eigenvalue $\lambda(p)$
of the Lax operator $\phi$ at the point $p$  is most easily
computed using the fact that $T(z) = \lambda(z)^{2}$.
Hence, $\lambda_{l} = \sum_{i} {{Z_{i}}\over{p_{l} - z_{i}}}$,
\be
 X_{i}^{+} = u {{P(z_{i})}\over{Q^{\prime}(z_{i})}}, \\
 Z_{i} = \sum_{l} {{\lambda_{l}}\over{z_{i} - p_{l}}}
{{Q(p_{l})P(z_{i})}\over{Q^{\prime}(z_{i})P^{\prime}(p_{l})}}
\ee
\beq
 \qquad P(z) = \prod_{l=1}^{k-1} ( z - p_{l})
 \qquad Q(z) = \prod_{i=1}^{k} (z - z_{i})
\eeq
The value of $u$ can be set to $1$ by the $\IC^{*}$ transformation.
The $({\lambda_{l}}, p_{l})$'s are
the gauge invariant coordinates on $\CP_{0}$. They are
defined up to a permutation.
It is easy to check that the restriction of the symplectic form
$\sum \omega_{i}$ onto the set $\sum_{i} Z_{i} = 0$
is the pullback of the form
\beq
\sum_{l=1}^{k-1} d \lambda_{l} \wedge d p_{l}.
\eeq

\subsection{Points on ${\IC} \times {\IC}$}
In this section we study the Hilbert scheme of points on
$S = {\IC}^2$, $S = {\IC}^{2} /{\Gamma}$ for ${\Gamma} \approx {\IZ}_{N},
{\IZ}$. We show that $S^{[v]}$ has a complex deformation $S^{[v]}_{\zeta}$
and that each $S^{[v]}_{\zeta}$ is an
integrable model including the complexification
of Calogero model .

Let us start with ${\IC}^2$. As  is well-known \cite{nakajima}
it is the set of
stable triples $(B_1, B_2, I)$, $I \in V \approx {\IC}^{v}, B_{1}, B_{2} \in
{\rm End}(V)$, $[B_{1}, B_{2} ] = 0$ modulo the action of ${\rm GL}(V)$:
$(B_{1}, B_{2}, I) \sim (g B_{1} g^{-1}, g B_{2} g^{-1}, g I)$ for
$g \in {\rm GL}(V)$.
Stability means that by acting on the vector $I$ by arbitrary polynomials
in $B_{1},B_{2}$ one can generate the whole of $V$.

The meaning of the vector $I$ and the operators $B_{1}, B_{2}$
is the following.
Let $z_{1}, z_{2}$ be the coordinates on ${\IC}^2$. Let $Z$ be
a zero-dimensional subscheme of ${\IC}^2$ of length $v$. It means that
the space ${\H}^{0}({\CO}_{Z})$
of functions on $Z$ which are the restrictions of holomorphic
functions on ${\IC}^2$ has dimension $v$. Let $V$ be this space of
functions. Then it has the canonical vector $I$ which is the constant
function $f = 1$ restricted to $Z$ and the natural action of
two commuting operators: multiplication by $z_1$ and by
$z_2$, which are represented by the operators $B_1$ and $B_2$.
Conversely, given a stable
triple $(B_{1}, B_{2}, I)$ the scheme $Z$, or, rather
the corresponding ideal ${\CI}_{Z} \subset {\IC} [ z_1, z_2 ]$
is reconstructed as follows: $f \in {\CI}_{Z}$ iff $f (B_{1}, B_{2} ) I = 0$.

Now let us discuss another aspect of the space $\left( {\IC}^{2} \right)^{[v]}$.
It is symplectic manifold. To see this let us start with the space
of quadruples, $(B_{1}, B_{2}, I, J)$ with $B_{1}, B_{2}, I$ as above
and $J \in V^{*}$. It is a symplectic manifold with the symplectic form
\beq
\Omega = {\Tr}\left[  {\delta} B_{1} \wedge {\delta} B_{2} +
\delta I \wedge {\delta} J  \right]
\eeq
which is invariant under the naive action of $G = {\rm GL}(V)$. The moment map
for this action is
\beq
\mu = [B_{1}, B_{2} ] + IJ \in {\rm Lie}G .
\eeq
Let us perform the Hamiltonian reduction, that is
take the zero level set of $\mu$, choose a subset of stable points in the
sense of GIT and take the quotient
of this subset with respect to $G$. One can show \cite{nakajima}
that the stability implies that $J =0$ and therefore the moment equation reduces
to the familiar $[B_1, B_2]=0$.

Moreover, $\left( {\IC}^{2} \right)^{[v]}$ is an integrable system. Indeed,
the functions ${\Tr} B_{1}^{l}$ Poisson-commute and are
functionally independent (in open dense subset) for $l= 1, \ldots, v$.

It turns out that $\left( {\IC}^{2} \right)^{[v]}$ has an interesting
complex deformation which preserves its symplecticity and
integrability. Namely, instead of $\mu^{-1}(0)$ in the reduction one should take
${\mu}^{-1} ( \zeta \cdot {\Id} )$ for some $\zeta \in {\IC}$.
Now $J \neq 0$. The resulting quotient $S_{\zeta}^{[v]}$ no longer parametrizes
subschemes in ${\IC}^2$ but rather sheaves on a non-commutative ${\IC}^{2}$,
that is the ``space'' where functions are polynomials in $z_1, z_2$ with the
commutation relation $[z_1, z_2] = \zeta$ \cite{ns} .
Nevertheless, the quotient itself is a perfectly well-defined symplectic
manifold with an integrable system on it:
the functions $H_{l}= {\Tr} B_{1}^{l}$
still Poisson-commute and are
functionally independent for $l= 1, \ldots, v$. On the dense open subset
of $S^{[v]}_{\zeta}$ where $B_{2}$ can be diagonalized:
$B_{2} = {\rm diag} \left( q_1, \ldots, q_{v} \right)$ the Hamiltonians
$H_{1}, H_{2}$ can be written as follows:
\beq
H_{1} = \sum_{i} p_{i}, \quad H_{2} = \sum_{i} p_{i}^{2} +
\sum_{i < j} {{\zeta^2}\over{\left( q_{i} - q_{j} \right)^2}}
\eeq
where $p_{i} = \left( B_{1} \right)_{ii}$. These Hamiltonians
describe a collection of indistinguishable particles on a (complex)
line with a pair-wise potential interaction $1\over{x^2}$. This system
is called rational Calogero model . It is shown in \cite{wilson} that
the space $S^{[v]}_{\zeta}$ can be used for compactifying the Calogero
flows in the complex case and moreover that the same compactification
is natural in the KdV/KP realization of Calogero flows .

\subsection{Generalization to ${\IC} \times {\IC}^{*}$ and to
${\IC}^{*} \times {\IC}^{*}$}

Now let $z_{1}, z_{2}$ be the coordinates on ${\IC} \times {\IC}^{*}$, i.e.
$z_{2} \neq 0$. Then the description of the previous section is still
valid except that $B_{2}$ must be invertible now. So in this case the Hilbert
scheme of points is obtained by a complex Hamiltonian reduction
from the space $T^{*}\left( G \times  V \right)$. .
The moment map in our
notations will be:
\beq
\mu = B_{2}^{-1} B_{1} B_{2} - B_{1} + IJ
\eeq
which corresponds to the symplectic form:
\beq
\Omega = \delta {\Tr} \left[
B_{1} B_{2}^{-1} \delta B_{2} + I \delta J \right]
\eeq
The reduction at the non-zero level $\mu = \zeta \cdot {\Id}$ leads to the
the complex analogue of either
Sutherland   or rational Ruijsenaars model . In the former case
$H_{2} = {\Tr} B_{1}^{2}$ while in the latter
$H_{rel} = {\Tr} \left( B_{2}  + B_{2}^{-1} \right)$. On the open dense
subset where  $B_{2}$ diagonalizable:
$B_{2} = {\rm diag} \left( \exp ( 2\pi i q_{1}) , \ldots, \exp ( 2\pi i q_{v} )
\right)$ the Hamiltonian $H_{2}$ equals:
\beq
H_{2} = \sum_{i} p_{i}^2 +
\sum_{i < j} {{\zeta^2}\over{{\rm sin}^{2}
\left( \pi\left( {q_{i} - q_{j}} \right)\right)}}
\eeq

Finally, if both $B_{1}$ and $B_{2}$ are invertible then
we get the Hilbert scheme of points on ${\IC}^{*} \times {\IC}^{*}$.
Its complex deformation is a bit more tricky, though. It turns out that
it can be obtained via {\it Poisson reduction} of
$G \times G \times V \times V$. The integrable system
one gets in this case is the trigonometric case of
Ruijsenaars model \cite{gnru}.

\subsection{ALE models}

Slightly generalizing the results of
\cite{nakajima} one may easily present
the finite-dimensional symplectic quotient construction of
the Hilbert scheme of points on $T^{*}{\IP}^{1}$.
Take $\CV = \IC^{v}$, $A = T^{*}{\Hom}(\CV , \CV)$,
$\tilde A = A \oplus A$ and $X = \tilde A \oplus T^{*}({\Hom}(\CV, \IC))$.
The space $X$ is acted on by the group $\CG = GL(V) \times GL(V)$.
The maximal compact subgroup $\CU$ of $\CG$ preserves the hyperkahler
structure of $X$.
The hyperkahler quotient of $X$ with respect to ${\CU}$ is
the Hilbert scheme of points
on $T^{*}\IP^{1}$ of length $v$.

This space is an integrable system. We shall prove it in  more
general setting. Namely, let $S$ be
the deformation of the orbifold $\IC^{2} / {\IZ}_{k}$, where
the generator $\omega = e^{{2\pi i}\over{k}}$ of $\IZ_{k}$ acts as
follows: $(z_{0}, z_{1}) \mapsto (\omega z_{0}, \omega^{-1} z_{1})$.
The space $S^{[v]}$ can be described as a hyperkahler quotient.
Let us take $k+1$ copy of the space $\IC^{v}$, and denote the $i$'th
vector space as $V_{i}$, $i = 0, \ldots, k$. Let us consider the
space
$$
X = \bigoplus_{i=0}^{k} {\rm Hom}(V_{i}, V_{i+1}) \oplus
{\rm Hom}(V_{i+1}, V_{i}) \bigoplus {\rm Hom}( W, V_{0}) \oplus
{\rm Hom} (V_{0}, W),
$$
where $k+1 \equiv 0$.
The space $X$ has a natural hyperkahler structure, in particular
it has a holomorphic symplectic form:
\beq
\omega = {\Tr} \delta I \wedge \delta J + \sum_{i=0}^{k}
{\Tr} \delta B_{i,i+1} \wedge
\delta B_{i+1, i}
\eeq
where $B_{i,j} \in {\rm Hom}(V_{j}, V_{i})$, $I \in {\rm Hom} (W, V_{0})$,
$J \in {\rm Hom} (V_{0}, W)$.
The space $X$ has a natural symmetry group $G = \prod_{i=0}^{k} U(V_{i})$
which  acts on $X$ as:
$$
B_{i,j} \mapsto g_{i} B_{i,j} g_{j}^{-1}, \quad I \mapsto g_{0} I,
\quad J \mapsto J g_{0}^{-1}
$$
The action of the group $G$ preserves the hyperkahler structure of $X$.
The complex moment map has the form:
\beq
\mu_{i} = B_{i,i+1} B_{i+1,i} - B_{i,i-1} B_{i-1,i} +
\delta_{i,0}
IJ
\eeq
The space $S^{[v]}$ is defined as a (projective) quotient of
$\mu^{-1}(0)$ by the
action of the complexified group $G$, which we denote
as $G_{c}$.
There is a deformation $S^{[v]}_{\zeta}$ which depends
on $k$ complex
parameters $\zeta_{0}, \ldots, \zeta_{k}, \sum_{i} \zeta_{i}=0$
defined as
\beq
S^{[v]}_{\zeta} = \cap_{i} \mu_{i}^{-1}(\zeta_{i} \Id) / G_{c}
\eeq
Now we present the complete set of Poisson-commuting functions on
$S^{[v]}$:
define the ``monodromy'':
\beq
\IB_{0} = B_{0,1} B_{1,2} \ldots B_{k, 0}
\eeq
which transforms under the action of $G$ in the adjoint representation.
The invariants
\beq
f_{l} = {\Tr}  \IB^{l}_{0}, \quad l = 1, \ldots, v
\eeq
clearly Poisson-commute on $X$, are gauge invariant and therefore descend
to the commuting functions on $S^{[v]}$ (and to $S^{[v]}_{\zeta}$ as well).
The functional independence is easily checked on the dense open
set where $S^{[v]}$ can be identified with the symmetric product of
$S$'s.

\subsection{ Integrable model of type $A_n$  }

We will  consider some particular examples of the ADE or
"quiver"systems which are corresponded to the case $G=Z_{n+1}$.
Below we use the explicit computations for which we are thankful
to A. Kotov.

 Let  $\eps=exp(\frac{2\pi i}{n+1})$ be the generators
of the group $Z_{n+1}$ acting on $\IC^2$ as $\eps: (q_1, q_2)\rt
(\eps q_1, \eps^{-1}q_2)$.

First we consider the simplest case $n=1$. The phase space is a
quotient $X=\mu^{-1}(\zeta)/G$, where $\mu$ is a momentum map of
the symplectic action of $G=GL(V_0)\times GL(V_1)$ on
$M=T^*Hom(V_0,V_1)\op T^*Hom(V_1,V_0)\op T^*V_0$. One can choose a
local canonical system of coordinates in two ways. The first deals
with the eigenvalues of the monodromy matrix $B_0=B_{01}B_{10}$.
On the dense open subset of $X$ the eigenvalues $q_i$ are
different and nonvanishing such that $q_i=e^{\phi_i}$, $\phi_i\ne
\phi_j$ for $i\ne j$. Therefore it is possible to diagonalize
$B_0$ by $GL(V_0)$ action. Then using $GL(V_1)$ transformations we
obtain $B_{01}=Q$, where $Q=diag(q_i)$ and $B_{10}=I$. Notice that
the symmetry group which does not change the above mentioned
condition is the diagonal subgroup $G_0=\{(g,g)\in GL(V_0)\times
GL(V_1)\}$.

Now the momentum equation
$$
\begin{array}{cc}
A_{01}B_{10}-B_{01}A_{10}+IJ &= \zeta_0\\
A_{10}B_{01}-B_{10}A_{01} &= \zeta_1
\end{array}
$$
has the form
$$
\begin{array}{cc}
A_{01}-Q A_{10}+IJ &= \zeta_0\\
A_{10} Q- A_{01} &= \zeta_1.
\end{array}
$$
The solution of the equation (up to a conjugation by
the elements of $G_0$) is
$$
I_i=\zeta, \; J_i=1, \; (A_{10})_{ii}=\frac{p_i}{q_i},\;
(A_{10})_{ij}=\frac{\zeta q_i}{q_i-q_j}, i\ne j,
$$
where $\zeta=\zeta_0 +\zeta_1$. The variables $(p_i,\phi_i)$ are
canonical so $\om=\sum_{i} dp_i\we d\phi_i$.

Taking the matrix $S=B_{01}A_{10}$ we see that the traces $H_k=tr
S^k$ are in involution. This integrable system coincides with the
Sutherland model. Namely the function $H_2=\sum_i p_i^2
-\sum_{i<j}\frac{8\zeta^2}{sh^2( \frac{\phi_i-\phi_j}{2})}$ is the
Sutherland hamiltonian.

In other hand, we can suppose that the eigenvalues of
$S=B_{01}A_{10}$ are different and non-zero, then  $B$ might be
diagonalized by the $GL(V_0)-$action. Taking  $B_{01}=I$ and
$A_{01}=P$, $P=diag(p_i)$ we see  that the the momentum equations
become
$$
\begin{array}{cc}
A_{01}B_{10}-P+IJ &= \zeta_0\\
P-B_{10}A_{01} &= \zeta_1
\end{array}
$$
and can be equivalently expressed as
$$
\begin{array}{cc}
B_{10}^{-1} P B_{10}-P+ IJ &= \zeta\\
A_{01} &= B_{10}^{-1}(P-\zeta_1)
\end{array}.
$$
The symplectic form $\om=dtrA_{01}\we dB_{10}$ is written as
$$
\om = d tr (P-\zeta_1)dB_{10}B_{10}^{-1}.
$$
It can be shown (using the arguments of \cite{Nekra})that the
answer leads to the rational Ruijsenaars-Schneider model. Indeed,
we have
$$
(A_{01})_{ij}=-\frac{\zeta e^{-\sqrt{-1}\phi_i}}{p_j+\zeta -p_i}
\sqrt{\frac{P(p_i-\zeta)P(p_j+\zeta)}{-\zeta^2 P'(p_i)P'(p_j)}},
$$
where $P(x)=(x-p_1)\ldots (x-p_N)$ and $(p_i,\phi_i)$ are canonical
variables.
It is clear that $H_k=tr A_0^k$ are Poisson commuting integrals and
$tr(A_0 +A_0^{-1})$ is a Ruijsenaars Hamiltonian.

Now it is almost straightforward to study the
$A_n$ - generalization of the model:
we can introduce two different canonical systems of coordinates
on the momentum solution manifold
$$
\begin{array}{ccc}
A_{01}A_{10}-A_{0n}A_{n0}+IJ &= \zeta_0,\\
A_{12}A_{21}-A_{10}A_{01} &= \zeta_1,\\
\ldots\\
A_{n0}A_{0n}-A_{n n-1}A_{n-1 n} &= \zeta_n
\end{array}
$$
modulo the transformations from $G$.

Let us diagonalize  as above  the monodromy matrix
$A_0=A_{01}\ldots A_{n0}$ such that on the open dense subset
$A_{01}=Q , Q=diag(q_i), q_i=e^{\phi_i}$ and $\phi_i\ne\phi_j$ for
$i\ne j$. We can eliminate  other factors by $GL(V_1),\ldots,
GL(V_n)$ action that means $A_{12}=A_{23}=\ldots =A_{no}=I$. We
are using again the diagonal symmetry group
$G_0=\{(g,g,\ldots,g)\in GL(V_0)\times\ldots\times GL(V_n)\}$.
Therefore we obtain
$$
\begin{array}{ccccc}
Q A_{10}& - A_{0n}+IJ &= \zeta_0,\\
A_{21}& - A_{10} Q &= \zeta_1,\\
A_{32}& - A_{21}&= \zeta_2,\\
\ldots\\
A_{0n}& - A_{n n-1} &= \zeta_n
\end{array}.
$$
and the result reads
$(A_{10})_{ii}=\frac{p_i}{q_i}$, $(A_{10})_{ij}=\frac{\zeta q_i}{q_i -q_j}$.
Thus the set of Hamiltonians $H_k=tr (A_{01}A_{10})^k$ yield the
Sutherland systems.
In particular, we have
$H_2=\sum_i p_i^2 -\sum_{i<j}\frac{8\zeta^2}{sh^2(
\frac{\phi_i-\phi_j}{2})}$.

The next option is to diagonalize $A_{01}A_{10}$ such that
$A_{01}=P$, $A_{10}=A_{21}=\ldots=A_{n n-1}=I$.
The momentum equations are replaced to
$$
\begin{array}{ccccc}
P-A_{0n}A_{n0}+IJ &= \zeta_0,\\
A_{12}-P &= \zeta_1,\\
A_{32}-A_{12} &= \zeta_2,\\
\ldots\\
A_{n0}A_{0n}-A_{n-1 n} &= \zeta_n
\end{array}.
$$
yielding  the following system:
$$
\begin{array}{ccc}
P-A_{0n} P A_{0n}^{-1} +IJ &= \zeta_0,\\
A_{0n} &=(P+\zeta-\zeta_0)A_{0n}^{-1},\\
A_{12} &= A_{n0}A_{0n} -(\zeta_2 +\ldots +\zeta_n),\\
\ldots\\
A_{n-1 n} &= A_{n0}A_{0n} - \zeta_n.
\end{array}
$$
The symplectic form $\om=tr dA_{0n}\we dA_{n0}$ can be written as
$\om=-d tr (P+\zeta-\zeta_0)A_{0n}^{-1} d A_{0n}$. Once again we
obtain that the reduced system is the Ruijsenaars model in the
same line as for the case of the "deformation" of
$(\mathbb{C}^2/{\mathbb {Z}_2)}^{[N]}$.

\subsection{Double elliptic system}

In this section we shall present new results concerning the
identification of the proper phase space for the double elliptic
system. We shall argue that two forms of Hamiltonians of the
double elliptic system \cite{bmmm,fgnr} are equivalent. We follow
the idea that two-body problems of Calogero-Ruijsenaars type can
be attributed to the finite dimensional groups. Along this
approach instead of generalization to one loop and two-loop
algebras we use more complicated finite-dimensional groups. For
example trigonometric Ruijsenaars model can be associated with the
quantum group while the elliptic model with the Sklyanin algebra
\cite{sklyanin}. In the latter case Hamiltonian exactly coincides
with the generator of the Sklyanin algebra \cite{kriza}.

To derive double elliptic system we need more general Poisson
bracket algebra intrinsically attributed to four dimensional
manifold. We have found that the proper object is the
Mukai-Odesskii algebra on the intersection of the quadrics
\cite{mukai,odesskii}. The Mukai-Odesskii algebra provides a
(possibly singular) Poisson structure on the intersection of $n$
quadrics $Q_i$ in $\mathbb{C}P^{n+2}$. The case $n=2$ corresponds
to the Sklyanin algebra. The Poisson bracket for the coordinates
looks as follows \cite{odesskii}

\beq \{x_i,x_j\}=(-1)^{i+j}det(\frac{\partial Q_k}{\partial
x_l}),l \neq i,j,
\eeq

where $x_i$ are homogeneous coordinates in $\mathbb{C}P^{n+2}$.
The corresponding four dimensional manifold is "noncommutative"
(in Poisson sense).

We choose the following system of four quadrics in$\mathbb{C}^6$
which provides the phase space for two-body double elliptic system
\be
x_1^2 - x_2^2 =1 \\
x_1^2 - x_3^2 =k^2 \\
g^2x_1^2 + x_4^2 + x_5^2=1 \\
g^2 \tilde {k}^2 x_1^2 + \tilde {k}^2 x_4^2 +  x_6^2=1
\ee

The first pair of the equations yields the "affinization" of
projective embedding of the elliptic curve
into $\mathbb{C}P^{3}$ and the second pair provides the elliptic curve  which
locally is "fibered" over the first elliptic curve. If the coupling
constant $g$ vanishes the system is just  the  two copies
of an elliptic curve embedded in $\mathbb{C}P^{3}\times\mathbb{C}P^{3}$.
Let us emphasize that the coupling constant
amounts to the additional noncommutativity between the
coordinates compared to the standard noncommutativity of coordinates
and momenta.

The above system of the four quadrics can be considered as an affine
part of a singular intersection surface $S \in \mathbb{C}P^{6}$:
\be
x_1^2 - x_2^2 = x_0^2\\
x_1^2 - x_3^2 =k^2 x_0^2 \\
g^2x_1^2 + x_4^2 + x_5^2=x_0^2 \\
g^2 \tilde {k}^2 x_1^2 + \tilde {k}^2 x_4^2 +  x_6^2=x_0^2.
\ee
This is a complex singular surface i.e. two-dimensional variety.

Taking the
point $(0;0;0;1;0;0)$ outside the surface $S$ one can consider a
"stereographic projection" on $\mathbb{C}P^5$ such that the surface is a
double-ramified covering of the following "singular $K3$-surface "
in $\mathbb{C}P^5$ given
by the system of three quadrics:
\be
x_1^2 - x_2^2 =1 \\
x_1^2 - x_3^2 =k^2 \\
\tilde {k}^2 (1 - x_5^2) +  x_6^2=1
\ee

The relevant Poisson brackets for this particular system of quadrics
read ( after the proper rescaling)
\be
\{x_1,x_2\}= \{x_1,x_3\}=\{x_2,x_3\}=0 \\
\{x_5,x_1\}= - x_2 x_3 x_4 x_6 \\
\{x_5,x_2\}= - x_1 x_3 x_4 x_6 \\
\{x_5,x_3\}= - x_1 x_2 x_4 x_6 \\
\{x_5,x_4\}=  g^2 x_1 x_2 x_3 x_6 \\
\{x_5,x_6\}=0
\ee
The nontrivial commutation relations between coordinates on the
distinct tori correspond to the standard phase space Poisson
brackets while the nontrivial bracket $\{x_5,x_4\}$
means the "additional noncommutativity" of the momentum space.
Let us note that the triple $x_1 x_2 x_3$ can be considered in the
elliptic parametrization
\be
x_1=\frac{1}{sn(q|k)}\\
x_2=\frac{cn(q|k)}{sn(q|k)}\\
x_3=\frac{dn(q|k)}{sn(q|k)}
\ee

Let us choose $x_5$ as the Hamiltonian of the double elliptic system.
The simple analysis shows that it is equivalent to the
Hamiltonian suggested in \cite{bmmm}. The dual hamiltonian can be
identified with ${x_2}\over{x_1}$.
When we fix a level of the Hamiltonian $E$ and the consistency condition
\beq
\tilde{k}^2 (1-E^2)=1-c^2
\eeq
one can reduce the number of the quadrics in the system because of the
coincidence of the last two equations. To get the explicit form of the
spectral curve
let us put $y = x_2^2x_3^2x_4^2$ and obtain the following
equation
$$
y^2 = (x_1^2 - 1)(x_1^2 - k^2 )(1-E^2 - g^2 x_1^2).
$$

If we take $z = x_1^2$ we obtain that the curve under consideration is
hyperelliptic of the genus 2
$$
y^2 = (1 - z)(k^2 - z)(1-E^2 - g^2 z).
$$
with "double" branching points $1, k^2,(1-E^2)/g^2$.
We can immediately use the curve for integration of the system:

$$
dt = {{dx_1}\over{x_2 x_3 x_4}} = {{dx_1}\over{y}},
$$

$$
t = \int{{dx_1}\over{\sqrt{(x_1^2 - 1)(x_1^2 - k^2 )(1-E^2 - g^2 x_1^2)}}}.
$$
More detailed derivation of the double elliptic systems from the
Mukai-Odesskii algebras shall be presented elsewhere \cite{bgr}.

Le us briefly consider degenerations. If one sends $\tilde k$ to
zero the system degenerates to the elliptic Ruijsenaars
model and the corresponding four dimensional manifold
reduces to $[T^2 \times \mathbb{C}^*]_{g}$ . At the next step one can
reduce the model to the elliptic Calogero model with
the manifold $[T^2 \times \mathbb{C}]_{g}$. Oppositely if $k$ is sent
to zero the system reduces to the one which is dual
to elliptic Ruijsenaars and Calogero models respectively.

The constructions involving quadrics provide very explicit
description of the noncommutative manifolds above. Indeed
let us consider limit $\tilde k$ which amounts to the
system of three quadrics in $\mathbb{C}P^5$
\be
x_1^2 - x_2^2 = 1 \\
x_1^2 - x_3^2 = k^2 \\
g^2 x_1^2 + x_4^2  + x_5^2  = 1.
\ee
This system corresponds to the K3 manifold which is the
phase space of the elliptic Ruijsenaars system. Geometrically
it looks as the cylinder bundled over the elliptic curve. The
general discussion on the completely integrable systems
associated with K3 manifolds can be found in \cite {taka}.

Let us discuss the relation with the Sklyanin algebra \cite{sklyanin}
visible in this respect geometrically. To this aim let
us interpret the Poisson bracket relations in the Sklyanin algebra
\beq
\{S_\alpha ,S_0\} = 2 J_{\alpha,\beta}S_{\beta} S_{\gamma}
\eeq
\beq
\{S_\alpha ,S_ \beta\} = 2 S_0 S_\gamma
\eeq
as example of the Mukai-Odesskii algebra generated by two
quadrics in $\mathbb{C}P^4$. The quadrics
\beq
K_1=\sum_{n=1}^{3} S_{n}^2 , K_2=S_{0}^2 +\sum_{n=1}^{3} J_n S_{n}^2
\eeq
coincide with the center of the Poisson bracket algebra. Hence
the Sklyanin aglebra fits with the general scheme.

Now let us remind the observation made in \cite{kriza} that the
Hamiltonian of the elliptic Ruijsenaars system coincides
with the generator of the Sklyanin algebra $S_0$. On the other
hand moving along our approach the elliptic Ruijsenaars
Hamiltonian coinsides with coordinate $x_5$. This means
that in the limit of vanishing $\tilde {k}$ the system
of four quadrics in $\mathbb{C}P^6$ effectively reduces to a system
of two quadrics in $\mathbb{C}P^2$ and two different Hamiltonians
for elliptic Ruijsenaars model
linear in coordinates can be identified. The degeneration
to the trigonometric Ruijsenaars model can be performed in the
similar way and the corresponding Hamiltonian
can be expressed in terms of the coordinates which
follow from the realiztion of $U_{q}(SL(2))$
in terms of the intersection of two quadrics in $\mathbb{C}P_4$.
Some generalization of this picture for the
trogonometric many-body system
can be found in \cite{hasegawa}.

Let us emphasize that it is clear from analysis above that the
discussed form of the double elliptic system  is not the general
one. Indeed one could simply consider more general form of the
system of quadrics to generate the multiparametric Poisson
bracket. As an example one could introduce the "momentum coupling
constant" $\tilde g$ and generate the system which is totally dual
with respect to the change of coordinate and momenta as well as
two coupling constants.

For completeness let us note that the periodic Toda
two-body system also can be described in terms of the
quadratic algebra. To this aim one can consider the
following Poisson algebra
\be
\{A_1,A_2\}=2A_3(4-A_2) \\
\{A_3,A_2\}=A_2 \\
\{A_1,A_3\}=A_1+A_3^2
\ee
Then the Hamiltonian of the Toda system is linear function
\beq
H_{Toda}=1/2(A_1+A_2)
\eeq
This picture can be considered as the further
degeneration from the elliptic Calogero model via Inozemtzev limit.

\subsection{Brane interpretation}
In the brane terms separation of variables can be
formulated  as reduction to a system of D0 branes
on some four dimensional manifold. It reminds
a reduction to a system of point-like instantons
on a (generically noncommutative \cite{ns}) four manifold.
One more essential point is that separated variables
amount to some explanation of the relation of periodic
Toda chain above   and monopole chains. Indeed, monopole
moduli space have the structure resembling the
one for the Toda chain in separated variables; both
of them are the Hilbert schemes of points on the
similar four manifolds.

The abovementioned constructions of the separation of variables
in integrable systems on moduli spaces of holomorphic bundles
with some additional structures can be described
as a symplectomorphism between  the moduli spaces of
the bundles (more precisely, torsion free sheaves)
having different Chern classes.

To be specific let us concentrate on the moduli space $\CM_{\vec
v}$ of stable torsion free coherent sheaves ${\CE}$ on $S$. Let
${\hat A}_{S} = 1 - [ {\rm pt} ] \in H^{*} (S, \mathbb{Z})$ be the
$A$-roof genus of $S$. The vector $\vec v = Ch ({\CE}) \sqrt{\hat
A_{S}} = (r; \vec w; d - r) \in {H}^{*}(S, \mathbb{Z}), \vec w \in
\Gamma^{3,19}$ corresponds to the sheaves
 with the Chern numbers:
\beq ch_{0} ({\CE})  = r \in {H}^{0}(S ; \mathbb{Z}) \eeq \beq
ch_{1}({\CE})  = \vec w \in {H}^{2} (S; \mathbb{Z}) \eeq \beq
ch_{2}({\CE}) = d \in {H}^{4}(S; \mathbb{Z}) \eeq Type $II A$
string theory compactified on $S$ has BPS states, corresponding to
the $Dp$-branes, with $p$ even, wrapping various supersymmetric
cycles in $S$, labelled by $\vec v \in {H}^{*}(S, \mathbb{Z})$.
The actual states correspond to the cohomology classes of the
moduli spaces $\CM_{\vec v}$ of the configurations of branes.  The
latter can be identified with the moduli spaces $\CM_{\vec v}$ of
appropriate sheaves.

The string theory, compactified on $S$ has moduli space
of vacua, which can be identified with
$$
\CM_{A} = O\left( {\Gamma}^{4,20} \right) \backslash O(4,20;
\mathbb{R}) / O(4;\mathbb{R}) \times O(20;\mathbb{R})
$$
where the arithmetic group $O({\Gamma}^{4,20})$ is the group of
discrete authomorphisms. It maps the states corresponding to
different $\vec v$ to each other. The only invariant of its action
is ${\vec v}^{2}$.

We have studied three realizations of an
integrable system.
The first one uses the non-abelian
gauge fields on the
curve $\Sigma$ imbedded into symplectic
surface $S$. Namely,  the phase space of the system is the
moduli space of stable pairs: $(\CE, \phi)$, where $\CE$
is rank $r$ vector bundle over $\Sigma$ of degree
$l$, while $\phi$
is the holomorphic section of
$\omega^{1}_{\Sigma} \otimes {\rm End}({\CE})$.
The second realization is the moduli space of pairs $(C, {\CL})$,
where $C$ is the curve (divisor) in $S$ which realizes
the homology class $r[\Sigma]$ and $\CL$ is the line bundle on $C$.
The third realization is the Hilbert scheme of points on $S$
of length $h$, where $h = {\half}{\rm dim}{\CM}$.

The equivalence of the first and the second realizations
corresponds to the physical statement that the bound
states of $N$ $D2$-branes wrapped around $\Sigma$ are represented
by a single $D2$-brane which wraps a holomorphic curve $C$
which
is an $N$-sheeted covering of the base curve $\Sigma$.
The equivalence of the second and the third descriptions
is natural to attribute to $T$-duality.

Let us mention that the separation of variables
above provides some insights on the Langlands duality
which involves spectrum of the Hitchin Hamiltonians.
The attempt to reformulate Langlands duality as a quantum
separation of variables has been successful for the
Gaudin system corresponding to the spherical case \cite{frenkel}.
The consideration in \cite{fgnr}  suggests that the  proper
classical version of the Langlands correspondence
is the transition to the Hilbert scheme of points
on four-dimensional manifold. This viewpoint
implies that quantum case can be considered as
correspondence   between
the eigenfunctions of the Hitchin Hamiltonians
and solutions to the Baxter equation in the separated
variables.

\section{Discussion}

We would like to conclude with some general remarks concerning to
the application of the dualities above to other moduli spaces. The
systems of the Hitchin type are closely related to the instantons
and monopole moduli on the manifolds with the compact dimensions.
Namely the Hitchin system on the torus without marked points can
be mapped by T -duality transformation along two dimensions onto
the the moduli space of instantons on $\mathbb{R}^2\times T^2$
\cite{kapustin}. When the marked point is added then  T-duality
transforms it to the noncommutativity of the manifold where
instantons are considered on and the Hilbert scheme on the
noncommutative manifold emerges. That is what we have described at
another language via the separations of variables proceedure.

In a similar manner the relation to the monopole moduli can be
formulated. The Hilbert schemes enters the description of the
monopole moduli spaces very naturally. It is well known that
moduli space of monopoles on $\mathbb{R}^4$ in commutative theory
is geometrically $Hilb \mathbb{C}\times \mathbb {C}^*$ which on
the Hitchin side corresponds to the system on a cylinder. Recently
it was argued that the Hitchin theory on the cylinder with marked
points is mapped via a chain of dualities into  the moduli space
of $SU(2)$ monopoles on $\mathbb{R}^3\times S^1$ \cite{cherkis}
with prescribed asymptotics.

Finally let us comment on six dimensional theory on NS5 branes
compactified on a three dimensional torus ${\bf T}^{3}$ down to
three dimensions. As was discussed extensively in \cite{ganor} in
case where two out of three radii of ${\bf T}^{3}$ are much
smaller then the third one ${\bf R}$ the effective three
dimensional theory is a sigma model with the target space ${\CX}$
being the hyper-kahler manifold (in particular, holomorphic
symplectic) which is a total space of algebraic integrable system.
The complex structure in which ${\CX}$ is the algebraic integrable
system is independent of the radius ${\bf R}$ while the K\"ahler
structure depends on $\bf R$ in such a way that the K\"ahler class
of the abelian fiber is proportional to $1/{\bf R}$. In the limit
of the large R theory is effectively four-dimensional and the
dualities between integrable systems have the dualities between
the Coulomb moduli spaces as the field theory counterparts. This
issue has been reviewed in \cite{gormir}.

We would like to thank H.Braden, V.Fock, N. Nekrasov and
A.Odesskii for the collaboration on these issues and A. Mironov,
A. Morozov A. Marshakov and M. Olshanetsky for the fruitful
discussions. Our special thanks to A.Kotov who kindly supplied us
with his computations. A.G. thanks University of Angers where the
part of the work has been done for the kind hospitality. The
research of A.G. was supported in part by grants INTAS-99-1705,
CRDF-RP2-2247 and grant for research CNRF 2000. V.R. is grateful
to the organized committee of the Kiev NATO Advanced Workshop for
invitation, to INTAS -99-1705 and to RFFI 2000.

\end{document}